\documentclass[aps,twocolumn,groupedaddress,prl,floatfix]{revtex4}
\usepackage{amssymb}
\usepackage{amsmath}
\usepackage{epsfig}

\begin{document}

\bibliographystyle{apsrev}
\preprint{}

\title{Photocurrent, Rectification, and Magnetic Field Symmetry of \\
Induced Current Through Quantum Dots}
\author{L.~DiCarlo, C.~M.~Marcus}
\affiliation{Department of Physics, Harvard University, Cambridge,
Massachusetts 02138}
\author{J.~S.~Harris,~Jr.}
\affiliation{Department of Electrical Engineering, Stanford
University, Stanford, California 94305}
\date{\today}

\begin{abstract}

We report mesoscopic dc current generation in an open chaotic
quantum dot with ac excitation applied to one of the
shape-defining gates. For excitation frequencies large compared to
the inverse dwell time of electrons in the dot (i.e., GHz), we
find mesoscopic fluctuations of induced current that are fully {\em
asymmetric} in the applied perpendicular magnetic field, as
predicted by recent theory. Conductance, measured simultaneously, is
found to be symmetric in field.  In the adiabatic (i.e., MHz)
regime, in contrast, the induced current is always symmetric in field,
suggesting its origin is mesoscopic rectification.

\end{abstract}
 \maketitle

The study of phase-coherent electron transport in systems with
rapidly time-varying potentials significantly extends the domain
of mesoscopic physics, and is likely to be important in quantum
information processing in the solid state, where gate operations
must be fast compared to decoherence rates. Three regimes of
mesoscopic transport may be identified: dc interference effects
with static potentials, including effects such as universal
conductance fluctuations (UCF) and weak localization; adiabatic
dynamic effects such as rectification and mesoscopic charge
pumping
\cite{Thouless83,Spivak96,Brouwer98,Switkes99,Shutenko00,Polianski01,Brouwer01};
and high-frequency phenomena such as photovoltaic effects
\cite
{Falko89,Bykov89,Bykov90,Giordano92a,Giordano92b,Bykov93,Giordano96,Giordano97}
and decoherence from a fluctuating electromagnetic environment
\cite{Giordano89,Huibers99,Kravtsov01,Lin02}. Recently, theory
connecting these regimes in quantum dots has appeared
\cite{VavDeph,VavOns,VavPV,Moskalets02}, with most predicted
effects remaining unexplored experimentally.

In this Letter, we compare dc currents induced by adiabatic and
nonadiabatic sinusoidal modulation of the voltage on one of the
confining gates of an open GaAs/AlGaAs quantum dot. Motivated by
recent theory \cite{Shutenko00,Brouwer01,VavPV}, we pay particular
attention to the symmetry of the induced dc current as a function
of perpendicular magnetic field, $B$. For adiabatic frequencies
($\omega \ll \tau_d^{-1}$, where $\tau_d$ is the inverse dwell
time of electrons in the dot) mesoscopic fluctuations (about a zero average) of induced
current are always found to be symmetric in $B$. As discussed in
\cite{SwitkesPhD,Brouwer01}, this suggests {\em
rectification}---due to coupling of the gate voltage to the
reservoirs combined with gate-dependent conductance---as the
principal source of induced current in this regime. On the other
hand, for gate voltage modulation at GHz frequencies ($\omega
\gtrsim \tau_d^{-1}$), the induced dc current may be either
predominantly symmetric or completely asymmetric in $B$, depending
on the particular frequency. We interpret these results as showing
competing mechanisms of induced mesoscopic current in the GHz
regime, with rectification producing a signal symmetric in $B$ and
photocurrent producing a signal fully asymmetric in $B$. Which
effect dominates depends strongly on the frequency of the
modulation. We thus establish field symmetry as an experimental
tool for separating these different physical effects. This allows
us to study, for instance, how rectification and photocurrent
separately depend on the amplitude of the applied ac modulation.

The quantum dot investigated was formed by electrostatic gates $90\,
\mathrm{nm}$ above the two-dimensional electron gas (2DEG) in a
GaAs/AlGaAs heterostructure, and has an area $A\sim0.7\,
\mu\mathrm{m}^2$, accounting for $\sim 50\, \mathrm{nm}$ depletion
at the gate edge. The bulk density of the 2DEG was $2\times
10^{11}\, \mathrm{cm}^{-2}$ and the mobility $1.4\times10^{5}\,
\mathrm{cm}^{2}/\mathrm{Vs}$, giving a bulk mean free path $\sim
1.5\,\mu\mathrm{m}$. Ohmic contact resistances were on the order
of $350\, \Omega$. Measurements are presented for the case of one
fully conducting (spin degenerate) mode per lead, which gives an
average dot conductance $\langle g \rangle\sim e^2/h$. Relevant
time scales include the dot crossing time $\tau_{\rm cross} \sim
\sqrt{A}/v_F \sim 4\, \mathrm{ps}$ and dwell time within the dot
$\tau_d= h/2\Delta\sim 0.2\, \mathrm{ns}$ for single-mode leads
($\Delta=2\pi\hbar^2/m^*A \sim10\, \mu\mathrm{eV}$ is the quantum
level spacing, $v_F$ is the Fermi velocity and $m^*$ the effective
electron mass).

The induced dc current through the dot was measured in a dilution
refrigerator via $300\, \Omega$ leads, while two semirigid $50\,
\Omega$ coaxial lines allowed ac excitation to be coupled from the
high frequency source (Wiltron 6769B) to either of two gates of
the dot over a range $10\, \mathrm{MHz}$ to $20\, \mathrm{GHz}$. A
room temperature bias tee enabled ac and dc voltage to be
simultaneously applied to the chosen gate. With coaxial lines
attached, the base electron temperature, where all measurements
were carried out, was $\sim 200\, \mathrm{mK}$, determined from
UCF amplitude \cite{Huibers99} and Coulomb blockade peak widths
\cite{LK97}.  The ac excitation was chopped at $153\, \mathrm{Hz}$
by a square pulse of variable duty cycle, $p$ (given as a
percentage). The induced current was amplified with an ac-coupled
Ithaco 1211 current amplifier (nominal input impedance $2\,
\mathrm{k}\Omega$) and lock-in detected. Conductance was
simultaneously measured by applying a $17\, \mathrm{Hz}$, $2\,
\mu\mathrm{V}_{\mathrm{rms}}$ sinusoidal voltage bias across the
dot. Output of the same current amplifier was then detected using
a second lock-in at $17\, \mathrm{Hz}$. The measured conductance
$g_p$ was found to be a simple weighted average of the conductance
with excitation on $(g_{\mathrm{on}})$ and off
$(g_{\mathrm{off}})$, $g_p \approx p g_{\mathrm{on}}+
(1-p)g_{\mathrm{off}}$, as seen in Fig.\ 1(d).

\begin{figure} \label{Figure1}
\includegraphics[width=3.2in]{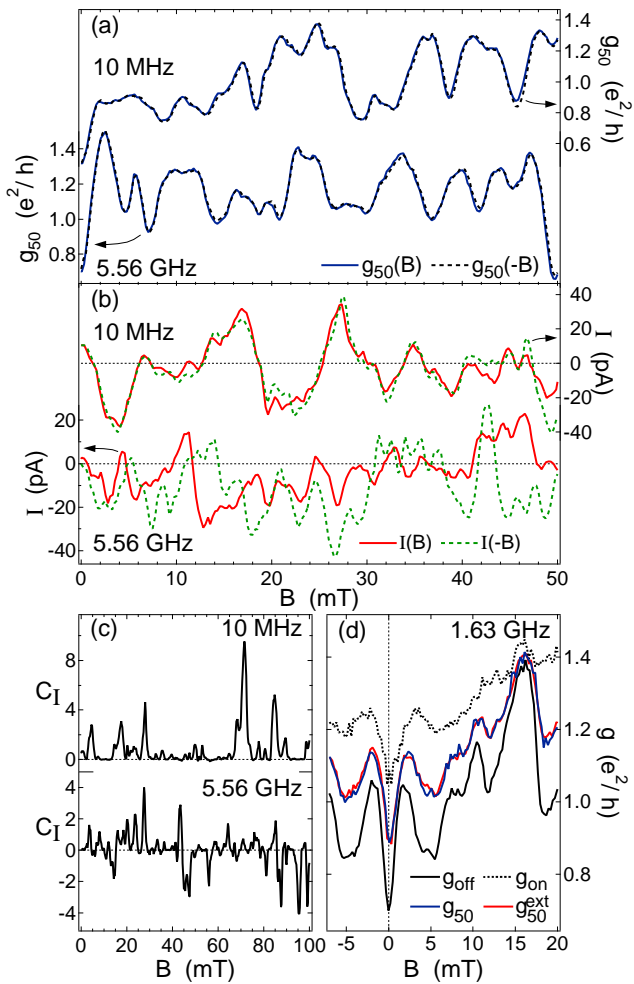} \caption{ \footnotesize{ (a)
Conductance $g_{50}(\pm B)$ as a function of perpendicular
magnetic field $B$, for $200\, \mathrm{nW}$ of incident radiation
at $10\, \mathrm{MHz}$ (top) and  $370\, \mathrm{nW}$ at $5.56\,
\mathrm{GHz}$ (bottom). (b) Induced dc currents $I(\pm B)$,
measured simultaneously with the conductance traces in (a). (c)
Cross-correlation $C_I(B)$ (see text) of induced current
fluctuations at $\pm B$.  Correlation is everywhere nonnegative
for $10\, \mathrm{MHz}$, but both positive and negative for
$5.56\, \mathrm{GHz}$. (d) Comparison of measured dot conductance
with 50\% duty cycle, $g_{50}$ (blue), and extracted trace
$g_{50}^{\rm ext} = (g_{\rm off} + g_{\rm on})/2$ (red), an
average of traces with ac excitation off $g_{\rm off}$ (black),
and excitation on, $g_{\rm on}$ (dashed). Traces at different
frequencies were made for different dot shape configurations, and
hence are uncorrelated. }}
\end{figure}

Fig.~1(a,b) shows dot conductance $g_{50}(\pm B)$ and induced
current $I(\pm B)$, with $p=50\%$ duty cycle, as a function of
perpendicular field, $B$, for applied frequencies of $10\,
\mathrm{MHz}$ $(\omega \tau_d\sim .01)$ and $5.56\, \mathrm{GHz}$
$(\omega \tau_d\sim 7)$. Dot conductances are in all cases found
to be symmetric in $B$, and show the expected weak localization
dip at $B=0$ and UCF. The induced current with $10\, \mathrm{MHz}$
excitation is also symmetric in field. In contrast, the induced
current for $5.56\, \mathrm{GHz}$ applied to the gate is found to
be fully asymmetric in field. This is evident in the traces in
Fig.\ 1(b). The difference between induced current at $10\,
\mathrm{MHz}$ and $5.56\, \mathrm{GHz}$ is seen most clearly by
looking at the correlation between induced currents at $\pm B$,
defined by $C_I(B)=(\delta I(B)\delta I(-B))/\langle\delta
I(B)^2\rangle_B$. The brackets denote an average over the measured
magnetic field range, and $\delta I$ is the deviation of the
current from its average value over this range. For $10\,
\mathrm{MHz}$, $C_I$ is non-negative for all magnetic field
values, whereas for $5.56\,\mathrm{GHz}$ it changes sign numerous
times.

The correlation field scales of mesoscopic fluctuations of induced
current and conductance for $10\,\mathrm{MHz}$ and $5.56\,
\mathrm{GHz}$ are roughly equivalent, $\sim1.5\, \mathrm{mT}$, as
determined from slopes of log-power spectra \cite{Huibers98}.
Induced dc current at $10\, \mathrm{MHz}$ $(5.56\, \mathrm{GHz})$
have roughly zero average and rms fluctuation amplitude of $\sim
20\, (13)\, \mathrm{pA}$ corresponds to $\sim 12 \, (.015)$
electrons per cycle. These values are obtained for ac gate
voltages of $6.4 (\sim12) \,\mathrm{mV}_{\mathrm{rms}}$,
comparable to the measured dc gate voltage correlation length,
$\sim 10\, \mathrm{mV}$.  (The ac gate voltage at 10 MHz was
calibrated and found equal to the voltage applied at the top of
the cryostat; the ac gate voltage at 5.56 GHz could not be easily
calibrated and instead was estimated by locating the inflection of
the curve of induced dc current versus incident power, as seen for instance in Fig.~3(d), and
comparing to theory \cite{VavPV}.)

  Figure 2 shows fluctuations of conductance and induced current
as a function of the dc gate voltage, $V_g^{dc}$, on a shape
distorting gate of the dot---the same gate to which the ac
excitation is applied---at opposite magnetic field values, $\pm
50$ mT. The conductance traces $g(V_g^{dc})$ in Fig.~2(a) are
nearly identical at $\pm 50$ mT, for both $10\, \mathrm{MHz}$ and
$2.4\, \mathrm{GHz}$ excitations. Induced current fluctuations,
shown in Fig. 2(b), are symmetric at $10\, \mathrm{MHz}$, but are
not symmetric for $2.4\, \mathrm{GHz}$.

Using the {\em same} gate both to drive the dot at ac and change
dot shape as a slowly swept parameter allows the induced dc
current to be compared to a simple model of rectification
\cite{SwitkesPhD, Brouwer01} applicable in the adiabatic limit.
The model takes the measured $g(V_g^{dc})$ as input, and assumes
that the ac part of the total gate voltage,
$V_g(t)=V_g^{dc}+V_g^{ac}\sin(\omega t)$, couples to the source
and drain reservoirs of the dot, giving rise to a (possibly phase
shifted) drain-source voltage, $V_{ds}(t)=\alpha
V_g^{ac}\sin(\omega t + \phi)$, with $\alpha$ and $\phi$ as
parameters. The dc rectification current resulting from
$V_{ds}(t)$ and $g(V_g(t))$ is given by

\begin{equation}
\label{Irect}
I_{\mathrm{rect}}=\frac{\omega}{2\pi}\int_0^{2\pi/\omega}\alpha
V_g^{ac}\sin(\omega t+\phi)g(V_{g}(t))dt.
\end{equation}

For $V_g^{ac}$ much less than the gate-voltage correlation scale,
$I_{\mathrm{rect}}\approx\frac{\alpha}{2}\cos(\phi) (V_g^{ac})^2
\frac{dg}{d V_{g}}$ \cite{Brouwer01}. However,  for the data in
Fig.~2, $V_g^{ac}=6.8\, \mathrm{mV}(\sim8\, \mathrm{mV})$ for
$10\, \mathrm{MHz}$ ($2.4\, \mathrm{GHz}$), which is not much
smaller than the correlation voltage of 10~mV, so the full
integral, Eq.~(\ref{Irect}), is used to model rectification.
Figure~2(b) shows a comparison of measured currents and
rectification currents calculated from Eq.~(\ref{Irect}), using
the $+50\, \mathrm{mT}$ data from Fig.~1(a) as input and values
$\alpha =7(6) \times 10^{-4}$ and $\phi = 0(0)$ for $10\,
\mathrm{MHz}$ ($2.4\, \mathrm{GHz}$). For the $10\, \mathrm{MHz}$
data, the similarity between the model and the measured current
suggests that rectification adequately accounts for the induced
current. On the other hand, the induced dc current at $2.4\,
\mathrm{GHz}$ $(\omega \tau_d\sim 3)$ does not appear to be well
described by the rectification model. This was to be expected: the
lack of symmetry in field of the induced current already tells us
that a rectification model that takes symmetric conductance as its
input cannot describe the asymmetric current induced by this
higher applied frequency.

\begin{figure}
      \label{Figure2}
     \includegraphics[width=3.2in]{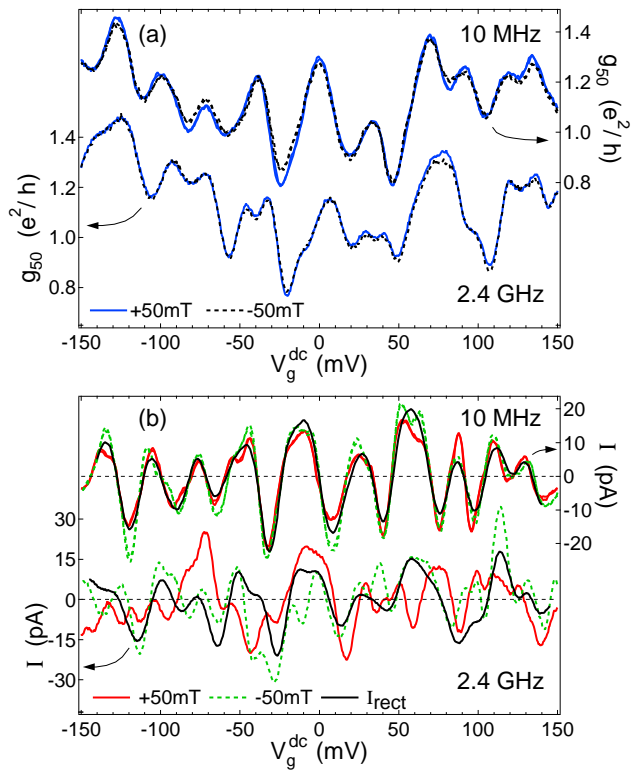}
\caption{ \footnotesize{ (a) Conductance as a function of the dc
voltage on the same gate to which ac is coupled, at $+50\,
\mathrm{mT}$ and $-50\, \mathrm{mT}$, for $115\, \mathrm{nW}$ of
incident power at $10\, \mathrm{MHz}$ (top)  and  $45\,
\mathrm{nW}$ at $2.4\, \mathrm{GHz}$ (bottom). Two traces are
shown at each magnetic field for $10\, \mathrm{MHz}$, revealing
the degree of repeatability in the measurements. Traces at
different frequencies were taken on different days. (b)
Simultaneously measurements of induced dc currents. Currents
calculated using the rectification model are also shown, in black,
for numerical parameters given in the text. }}
\end{figure}

\begin{figure}
      \label{Figure3}
     \includegraphics[width=3.25in]{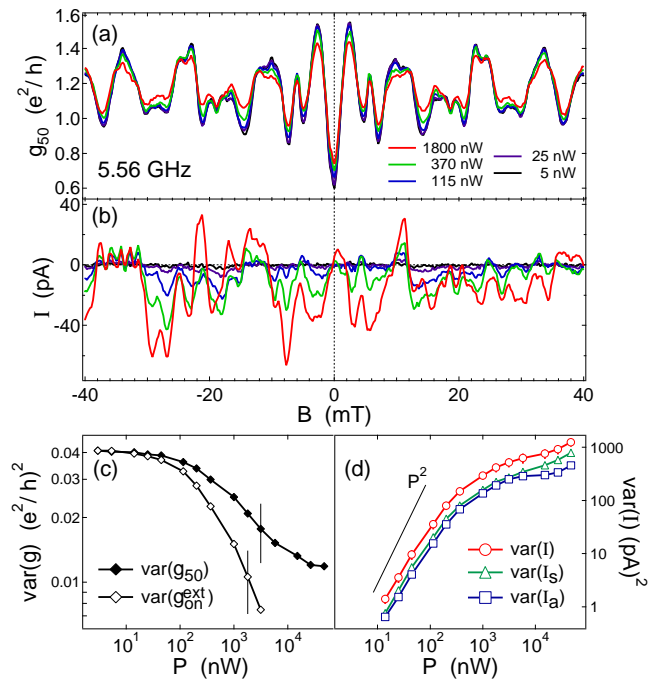}
\caption{\footnotesize{ Magnetic field dependence of (a)
conductance and (b) rf-induced dc current, for increasing levels
of incident power  $P$ at $5.56\, \mathrm{GHz}$. Full sweeps
spanned the $-100\, \mathrm{mT}$  to $+100\, \mathrm{mT}$ range.
(c) Variance of the measured conductance $g_{50}$ and of the
extracted on conductance $g_{\mathrm{on}}$, as a function of
incident power. (d) Power dependence of the total induced dc
current, and of its symmetric and antisymmetric components. }}
\end{figure}

Conductance and induced current as a function of perpendicular
field are shown in Fig.~3(a,b) for different levels of incident
power $P$  at $5.56\, \mathrm{GHz}$. Fluctuations in $g_{50}$
decrease with increasing $P$. In Figure 3(c), we show the power
dependence of the variance of the conductance. For high applied
powers, the variance of conductance fluctuations is reduced to
$1/4$ of its zero-radiation level. This is expected, since during
the fraction of time ($p=50\%$) that the radiation is on, UCF
should be fully suppressed at high power by a combination of shape
averaging, dephasing without heating, and heating effects
\cite{Kravtsov01,VavDeph}. Also shown in figure 3(c) is the
variance of the extracted ``on conductance,"
$g_{\mathrm{on}}^{\mathrm{ext}}=2g_{50}-g_{\mathrm{off}}$, which
we observe to decrease at a rate intermediate between  $P^{-1/2}$
and $P^{-1}$ at high power. Theory \cite{VavOns} predicts a rate
$P^{-1/2}$ in the absence of heating effects.

It is evident in Fig.~3(a) that no significant field asymmetry in
conductance is observed, regardless of ac power applied. Recent
theory \cite{VavOns} predicts that when applied frequency exceeds
the temperature in the leads (which is the case here, $k_B T_e/h
\sim 4\, \mathrm{GHz})$, dephasing will lead to field asymmetry in
conductance. However, for our experimental conditions the
predicted asymmetry is extremely small, smaller in fact than the
asymmetry due to drift and noise in our setup (which, as seen in
Figs.~1 and 2 is rather small.)

Figure 3(b) shows that mesoscopic fluctuations of the induced
current at $5.56\, \mathrm{GHz}$ are fully asymmetric in field and
increase in amplitude with increasing power.  As a measure of this
asymmetry, we compare the symmetric and antisymmetric components
of current, $I_{s(a)}(B)=(I(B)\pm I(-B))/2$.
Figure 3(d) shows their variances and that of total induced current,
calculated after subtracting a first order polynomial to $I(B)$ to
account for a non-mesoscopic background present at the highest
applied powers.

Variances of $I_s$ and $I_a$ are nearly equal, indicating that the
current is fully asymmetric for all powers (except at the highest
powers, discussed below). Variances of $I$, $I_s$, and $I_a$
increase at a rate approaching $P^{2}$ at the lowest powers,
consistent with theory \cite{VavPV} for the weak pumping limit.
The rate weakens at higher incident power, with a crossover near
$300\, \mathrm{nW}$, corresponding to an ac gate voltage of about
one correlation length.

\begin{figure} [bt]
      \label{Figure4}
      \includegraphics[width=3.2in]{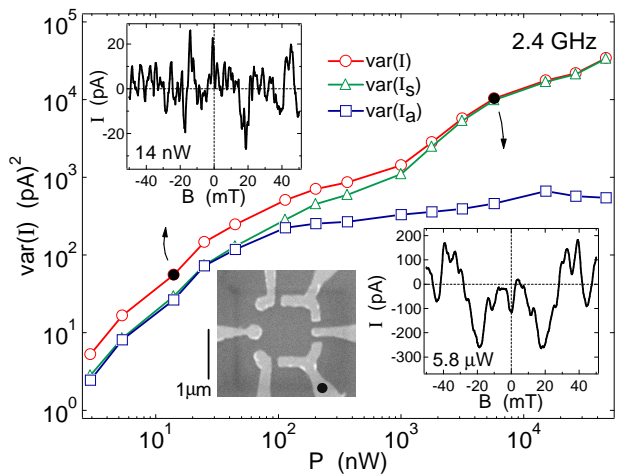}
\caption{ \footnotesize{Variance of symmetric, antisymmetric and
total current ($I_s$, $I_a$, $I$) as functions of incident power
$P$ at $2.4\, \mathrm{GHz}$. Upper left inset: Induced current as
a function of magnetic field, $B$, for $14\, \mathrm{nW}$ power,
showing asymmetric mesoscopic fluctuations with a correlation
length $\sim 1\,\mathrm{mT}$. Lower right inset: Induced current
as a function of magnetic field at $5.8\, \mu\mathrm{W}$ power,
showing symmetric fluctuations with a correlation length $\sim
4\,\mathrm{mT}$. An overall background in this trace was
subtracted with a first order polynomial $a_o+a_1B$. Lower left
inset: Micrograph of device. Dot indicates gate to which ac is
applied.}}
\end{figure}

In the nonadiabatic regime, the induced current is found to range
between being predominantly symmetric and fully asymmetric in $B$,
depending on the applied frequency. The degree of symmetry is also
found to depend on the applied ac power, with greater symmetry
found at higher power. This is illustrated in Fig.~4 for the case
of ac gate voltage at $2.4\,$GHz.  At low incident power the
current is fully asymmetric, and the fluctuations have a
correlation length $\sim 1\, \mathrm{mT}$. This is clearly
observed in the top left inset, where we show the magnetic field
dependence of the induced current for an incident power of $14\,
\mathrm{nW}$. At high power, $var(I_a)$ saturates, leaving a
predominantly symmetric signal (Fig.~4, lower inset). The
magnetic-field correlation length increases to $\sim 4\,
\mathrm{mT}$ at high power, suggesting enhanced dephasing
presumably due to the ac voltages on the gate causing electron
heating.

In summary, we have shown that field symmetry can be used to
distinguish mechanisms of induced dc current in response to ac
gate voltages in a quantum dot.  For lower frequencies, the
induced current is symmetric in field. This and other indications
suggest that mesoscopic rectification is responsible for the
induced current in this regime. For higher frequencies (GHz), the
induced current can be fully asymmetric in field, suggesting
instead that photocurrent can be the dominant source, as proposed
in recent theory \cite{Shutenko00,VavPV}.

We thank I.~Aleiner, P.~Brouwer, V.~Falko, A. Johnson, and
particularly M.~Vavilov for valuable discussion, and thank
C.~I.~Duru\"oz, S.~M.~Patel, and S.~W.~Watson for experimental
contributions. We acknowledge support from the NSF under
DMR-0072777.

 }
\end{document}